\documentclass[
 aps, pra,
 amsmath,amssymb,
 12pt,
 final,
tightenlines,
 nofootinbib,
 superscriptaddress,
 ]
{revtex4}

\usepackage[utf8x]{inputenc}
\usepackage[english]{babel}
\usepackage{graphicx,times}
\usepackage{natbib}

\def\squareforqed{\hbox{\rlap{$\sqcap$}$\sqcup$}}

\def\sq{\ifmmode\squareforqed\else{\unskip\nobreak\hfil
\penalty50\hskip1em\null\nobreak\hfil\squareforqed
\parfillskip=0pt\finalhyphendemerits=0\endgraf}\fi}

\def\utw{\smash{\rlap{\lower5pt\hbox{$\sim$}}}}

\def\udtw{\smash{\rlap{\lower6pt\hbox{$\approx$}}}}

\def\diameter{{\ifmmode\mathchoice
{\ooalign{\hfil\hbox{$\displaystyle/$}\hfil\crcr
{\hbox{$\displaystyle\mathchar"20D$}}}}
{\ooalign{\hfil\hbox{$\textstyle/$}\hfil\crcr
{\hbox{$\textstyle\mathchar"20D$}}}}
{\ooalign{\hfil\hbox{$\scriptstyle/$}\hfil\crcr
{\hbox{$\scriptstyle\mathchar"20D$}}}}
{\ooalign{\hfil\hbox{$\scriptscriptstyle/$}\hfil\crcr
{\hbox{$\scriptscriptstyle\mathchar"20D$}}}}
\else{\ooalign{\hfil/\hfil\crcr\mathhexbox20D}}%
\fi}}

\input{maik.rty}

\begin{document}

\title{Long-term high-resolution spectroscopic monitoring of the evolved metal-poor binary PS\,Gem}

\author{V.G.~Klochkova}
\affiliation{Special Astrophysical Observatory, Nizhnii Arkhyz, 369167 Russia}
\email{Valentina.R11@yandex.ru}
\author{V.E.~Panchuk}
\affiliation{Special Astrophysical Observatory, Nizhnii Arkhyz, 369167 Russia}
\author{J.K.~Zhao}
\affiliation{National Astronomical Observatories, Chinese Academy of Sciences,
             Beijing 100012, P.\ R.\ China}
\author{Y.Q.~Wu}
\affiliation{National Astronomical Observatories, Chinese Academy of Sciences,
             Beijing 100012, P.\ R.\ China}
\author{N.S.~Tavolzhanskaya}
\affiliation{Special Astrophysical Observatory, Nizhnii Arkhyz, 369167 Russia}
\author{G.~Zhao}
\affiliation{National Astronomical Observatories, Chinese Academy of Sciences,
             Beijing 100012, P.\ R.\ China}
\affiliation{School  of Space Science and Technology, University of Shandong,
             Shandong 264209, P.\ R.\ China}

\begin{abstract}
We present results of long-term high-resolution spectroscopic monitoring of the extremely low-metallicity star
PS\,Gem (=HD\,52961), associated with IRAS\,07008+1050. The spectra were obtained with the 6-m BTA telescope at
a resolving power of $R=60\,000$ during 1997–-2025, providing a homogeneous dataset spanning nearly three decades.
Quantitative measurements of the H$\alpha$ two-peaked profile reveal repeatable phase-dependent variations of
the peak-intensity ratio $I_{\rm V}/I_{\rm R}$ and the absorption-core velocity over multiple orbital cycles,
whereas the peak separation remains nearly constant at $\langle\Delta V_r\rangle = 78\pm4$\,km\,s$^{-1}$.
The heliocentric radial velocity measurements yield a systemic velocity of  $\gamma=7.46\pm 0.64$\,km\,s$^{-1}$
and an orbital semi-amplitude of $K=2.49$\,km\,s$^{-1}$. No significant long-term secular drift is observed over
the monitoring interval.  The stellar luminosity, estimated from the equivalent width of the O\,I~7773\,\AA{}
triplet ($W_{\lambda}=1.16\pm0.02$\,\AA{}), corresponds to $\log(L/L_{\odot}) \approx 3.5$. The average radial
velocity of diffuse interstellar bands is consistent with the location of PS\,Gem in the Local Arm.  These
observations provide rare multi-decade time-domain constraints on the dynamical behavior of this evolved   metal-poor star.
\end{abstract}
\maketitle

\section{Introduction}\label{sect:intro}

PS\,Gem (= HD\,52961 = IRAS\,07008+1050) is one of the most chemically peculiar evolved binary systems known in the Galaxy.
The star exhibits an extreme depletion pattern, with an iron abundance of approximately ${\rm [Fe/H]}\approx-4.8$, while
volatile elements such as C, N, O, S, and Zn remain comparatively abundant \cite{Waelkens91,Kipper}. This unusual chemical
composition is generally interpreted as the result of selective depletion associated with gas--dust separation in a circumbinary environment and subsequent reaccretion of metal-poor gas onto the stellar surface.

The star also displays a pronounced infrared excess, indicating the presence of circumstellar or circumbinary dust. Long-term
radial-velocity monitoring established that PS\,Gem is a binary system with an orbital period of $P=1310\pm8$\,d \cite{Winckel1999}, while later studies confirmed the existence of a stable circumbinary disk and further refined its atmospheric properties~\citep{Oomen}. Owing to its extreme depletion and relatively simple optical spectrum, PS\,Gem has become an important object for studying the interplay between binary evolution, circumbinary disks, and chemical depletion in evolved stars.

Despite extensive abundance studies, the long-term spectroscopic variability of PS\,Gem has received comparatively little attention. In particular, the variability of the complex H$\alpha$ profile and its relation to orbital motion have not been investigated using a homogeneous set of high-resolution spectra obtained over multiple orbital cycles. Such observations are essential for distinguishing orbital effects from atmospheric variability and for characterizing the circumstellar environment responsible for the observed emission features.

As part of a long-term spectral monitoring program of evolved stars carried out with the  6-meter BTA telescope of the Special Astrophysical Observatory, we obtained a homogeneous series of high-resolution spectra of PS\,Gem spanning nearly three decades (1997--2025). This dataset provides a unique opportunity to investigate the stability of the stellar atmosphere, the radial-velocity variability, and the long-term behaviour of the H$\alpha$ profile in a chemically peculiar evolved binary system.

In this paper, we present a quantitative analysis of the spectroscopic variability of PS\,Gem based on 16 high-resolution spectra. Particular attention is devoted to the orbital-phase dependence of the H$\alpha$ profile and its implications for the circumstellar environment of the system. Section~\ref{observ} describes the observations and data reduction procedures. Section~\ref{results}
presents the observational results. Section~\ref{discuss} discusses the origin of the observed variability and the evolutionary status of PS\,Gem, while the main conclusions are summarized in Section~\ref{conclusion}.

\section{Observations and Data Reduction}\label{observ}

The first spectrum of PS\,Gem was obtained in 1997 using the moderate-resolution ($R \approx 15\,000$) echelle spectrograph
PFES~\citep{PFES} at the prime focus of the 6-m BTA telescope. All subsequent observations were carried out with
the high-resolution echelle spectrograph NES~\citep{NES}, permanently mounted at the Nasmyth focus of the same telescope.
The NES spectra were obtained at a resolving power of $R \ge 60\,000$. The signal-to-noise ratio (S/N) of the spectra ranges from 50 to 110, depending on observing conditions and wavelength region. The observing dates and corresponding setups are summarized in Table~\ref{tab:obs_rv}.

The NES spectrograph is equipped with a $4608 \times 2048$ pixel CCD detector with a pixel size of 13.5\,$\mu$m and a readout
noise of 1.8\,e$^{-}$. In most observing runs, the spectral coverage spans 470--778\,nm. Minor variations in wavelength
coverage occurred during several runs due to detector replacement or modifications in the optical configuration.

To reduce slit losses while maintaining spectral resolution, NES employs a three-slice image slicer.
Data reduction was performed using a modified ECHELLE context~\citep{echelle} within the MIDAS software package,
adapted to the geometry of the NES echelle format. The reduction procedure included bias subtraction, flat-field
correction, order extraction, scattered light removal, and wavelength calibration using Th–Ar hollow-cathode lamp spectra. Cosmic-ray events were removed by median combining consecutive exposures obtained during the same night.
Further processing was carried out with the DECH20t package by \citet{dech}, which was used for continuum normalization,
measurement of line profiles, and radial velocity determinations.

All radial velocities were corrected to the heliocentric frame. Systematic uncertainties, estimated from telluric lines,
do not exceed 0.2\,km\,s$^{-1}$ for individual lines. The typical random uncertainty derived from numerous stellar absorption
lines is approximately 0.5\,km\,s$^{-1}$.
Line identification was performed using the spectral atlas of the post-AGB star CY\,CMi \citep{atlas}, observed with
the same BTA\,+\,NES configuration. The atlas provides a consistent reference for line identification over a wide wavelength
range under similar instrumental conditions.

\begin{table*}
\centering
\caption{Log of spectroscopic observations of PS\,Gem and heliocentric radial-velocity measurements.
Orbital phases were computed using  the orbital ephemeris of \citep{Winckel1999}.
Radial velocities were measured using  cross-correlation with selected photospheric absorption lines.
}
\label{tab:obs_rv}
\setlength{\tabcolsep}{3.5pt}
\begin{tabular}{cccccccc}
\hline
Date & JD-2450000 & Phase & Range &
$V_r({\rm abs})$ & $\sigma$ & $V_r({\rm H}\beta)$ & $V_r({\rm H}\alpha)$ \\
& & & (nm) & (km\,s$^{-1}$) & (km\,s$^{-1}$) & (km\,s$^{-1}$) & (km\,s$^{-1}$) \\
\hline
05.11.2008 & 4775.54 & 0.84 & 522--669 & $-5.19$ & 0.11 & $-2.70$ & --- \\
06.11.2009 & 5142.44 & 0.12 & 470--778 & 21.08 & 0.87 & --- & 12.35 \\
02.02.2010 & 5230.19 & 0.18 & 470--778 & 19.44 & 0.37 & 14.48 & --- \\
04.02.2010 & 5232.35 & 0.18 & 470--778 & 19.85 & 0.37 & --- & 4.64 \\
02.04.2010 & 5289.32 & 0.22 & 470--778 & 17.09 & 0.75 & --- & 5.15 \\
25.09.2010 & 5464.50 & 0.37 & 470--778 & 10.40 & 0.09 & --- & 3.82 \\
20.11.2010 & 5521.36 & 0.42 & 470--778 & 8.02 & 1.13 & 7.12 & --- \\
15.09.2011 & 5819.58 & 0.65 & 470--778 & 0.10 & 0.72 & $-0.74$ & $-2.04$ \\
27.11.2012 & 6258.55 & 0.99 & 470--778 & 10.08 & 0.59 & 11.95 & 10.14 \\
01.12.2012 & 6263.46 & 0.99 & 470--778 & 7.92 & 0.91 & 9.47 & 9.04 \\
13.02.2017 & 7796.51 & 0.18 & 470--778 & 21.36 & 0.58 & 19.48 & 9.34 \\
26.02.2018 & 8175.52 & 0.47 & 470--778 & $-0.14$ & 1.02 & $-1.83$ & $-5.60$ \\
05.04.2018 & 8214.22 & 0.51 & 470--778 & 1.04 & 1.28 & $-0.76$ & $-4.58$ \\
11.09.2022 & 9833.57 & 0.76 & 470--778 & $-3.09$ & 1.17 & $-2.48$ & $-0.01$ \\
06.11.2022 & 9889.51 & 0.81 & 470--778 & $-3.89$ & 0.81 & $-3.47$ & 1.71 \\
04.10.2025 & 10951.52 & 0.01 & 365--630 & $-4.66$ & 0.64 & $-5.72$ & --- \\
\hline
\end{tabular}
\end{table*}

\begin{table*}
\centering
\caption{Quantitative H$\alpha$ profile parameters of PS\,Gem. Columns list the observing date, orbital
phase, velocities of the blue and red emission peaks, their relative peak intensities
$I_{\rm V}$ and $I_{\rm R}$, peak separation $\Delta V_r$, radial velocity of the absorption core,
and equivalent width. All velocities are in km\,s$^{-1}$, and EW is given in m\AA.}
\label{tab:Halpha}
\setlength{\tabcolsep}{3.2pt}
\begin{tabular}{ccccccccc}
\hline
Date & Phase &  $V_r({\rm V})$ & $V_r({\rm R})$ &  $I_{\rm V}$ & $I_{\rm R}$ & $\Delta V_r$ & $V_r({\rm core})$ & EW \\
\hline
05.11.2008 & 0.84 & --- & --- & --- & --- & --- & --- & --- \\
06.11.2009 & 0.12 & $-41.90$ & 37.60 & 1.157 & 1.036 & 79.50 & 12.35 & 1224.69 \\
02.02.2010 & 0.18 & --- & --- & --- & --- & --- & --- & --- \\
04.02.2010 & 0.18 & --- & 73.84 & --- & 1.110 & --- & 4.64 & 1610.52 \\
02.04.2010 & 0.22 & --- & 83.66 & --- & 1.089 & --- & 5.15 & 1779.87 \\
25.09.2010 & 0.37 & --- & --- & --- & --- & --- & 3.82 & 2096.55 \\
20.11.2010 & 0.42 & --- & --- & --- & --- & --- & --- & --- \\
15.09.2011 & 0.65 & $-67.07$ & 9.25 & 1.100 & 1.280 & 76.32 & $-2.04$ & 1241.52 \\
27.11.2012 & 0.99 & $-41.29$ & 31.72 & 1.220 & 1.100 & 73.00 & 10.14 & 1286.25 \\
01.12.2012 & 0.99 & $-42.38$ & 32.48 & 1.190 & 1.060 & 74.86 & 9.04 & 1663.81 \\
13.02.2017 & 0.18 & $-0.91$ & 81.43 & 1.056 & 1.000 & 82.34 & 9.34 & 1743.30 \\
26.02.2018 & 0.47 & --- & 65.18 & --- & 1.000 & --- & $-5.60$ & 2347.93 \\
05.04.2018 & 0.51 & $-15.00$ & 70.30 & --- & 1.137 & 85.30 & $-4.58$ & 1565.16 \\
11.09.2022 & 0.76 & $-65.39$ & 8.80 & 1.128 & 1.100 & 74.19 & $-0.01$ & 1284.31 \\
06.11.2022 & 0.81 & $-67.16$ & 10.21 & 1.040 & 1.050 & 77.37 & 1.71 & 1291.96 \\
04.10.2025 & 0.01 & --- & --- & --- & --- & --- & --- & --- \\
\hline
\end{tabular}
\end{table*}

\section{Results}\label{results}

\subsection{Radial Velocity Variability and Atmospheric Stability}

The optical spectrum of PS\,Gem is dominated by absorption features and lacks detectable metallic emission lines. The only
prominent emission phenomenon is associated with the H$\alpha$ profile, whereas the H$\beta$ line remains purely absorptive
and is well reproduced by the theoretical profile calculated by \citet{Kipper}. No spectroscopic signatures attributable to
a secondary companion were detected in any of our spectra.

\begin{figure}[hbtp]
\includegraphics[angle=0,width=0.6\textwidth,bb=0 0 550 675,clip]{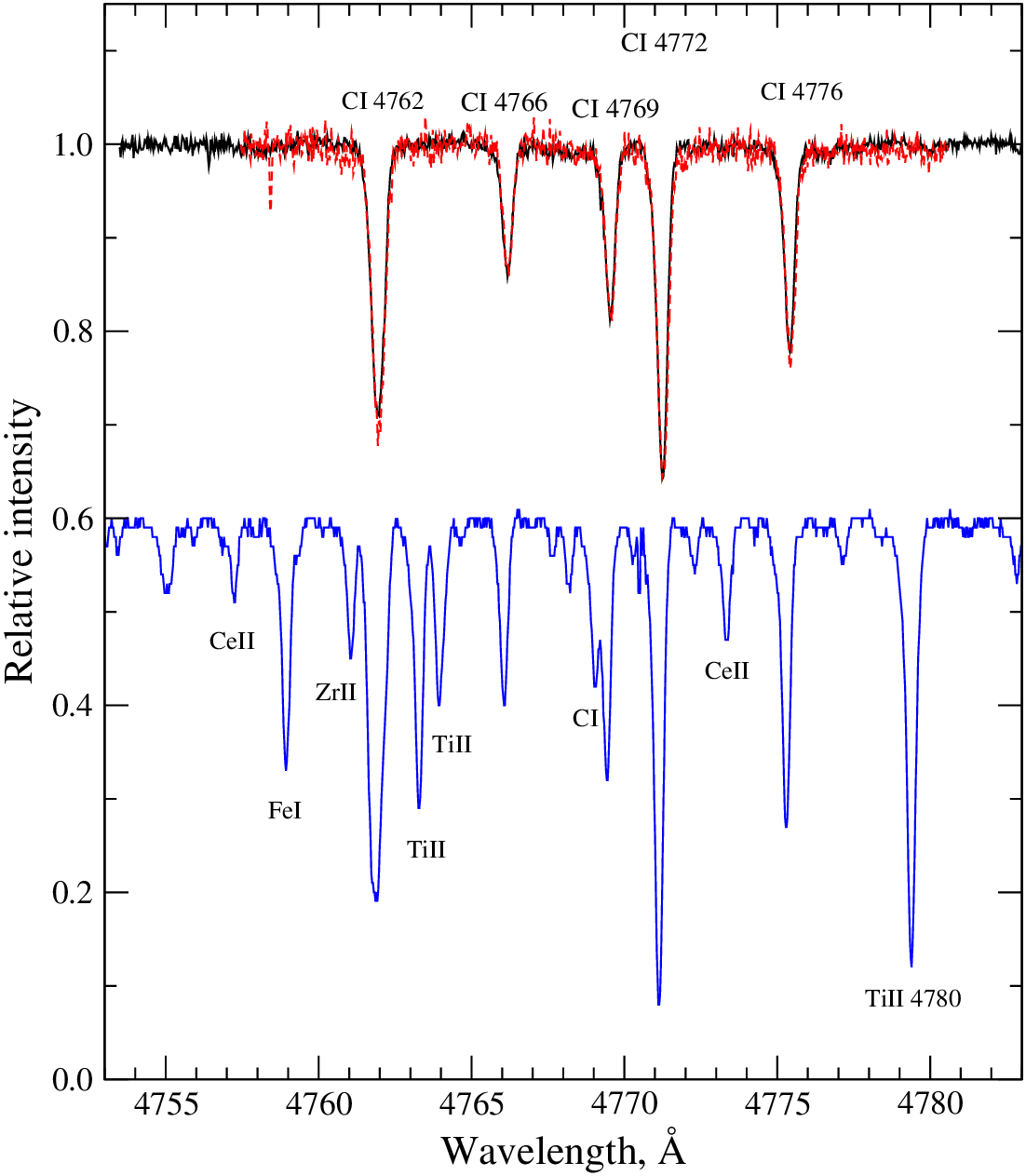} 
\caption{Fragments of the spectra of two post-AGB stars.
The upper panel shows two spectra of PS\,Gem obtained at orbital phases 0.47 (red dotted line) and 0.76 (solid black line), shifted relative to each other along the wavelength axis to compensate for the difference in radial velocity. The lower panel displays a fragment of the spectrum of CY\,CMi, shifted vertically by 0.4 for clarity.}
\label{CI_477}
\end{figure}

The spectrum is relatively poor in absorption features owing to the extremely low metallicity of the star. Absorption
lines of iron-group elements are almost absent, which complicates both spectral classification and radial-velocity measurements. Nevertheless, numerous lines of volatile elements, particularly C, O, Na, Mg, S, and Zn, are present. Neutral carbon lines
dominate the optical spectrum and provide the primary basis for radial-velocity determinations. Figure~\ref{CI_477} shows a
representative spectral region containing several C\,I lines.

\begin{figure}[hbtp]
\includegraphics[angle=0,width=0.5\textwidth,bb=0 0 550 675,clip]{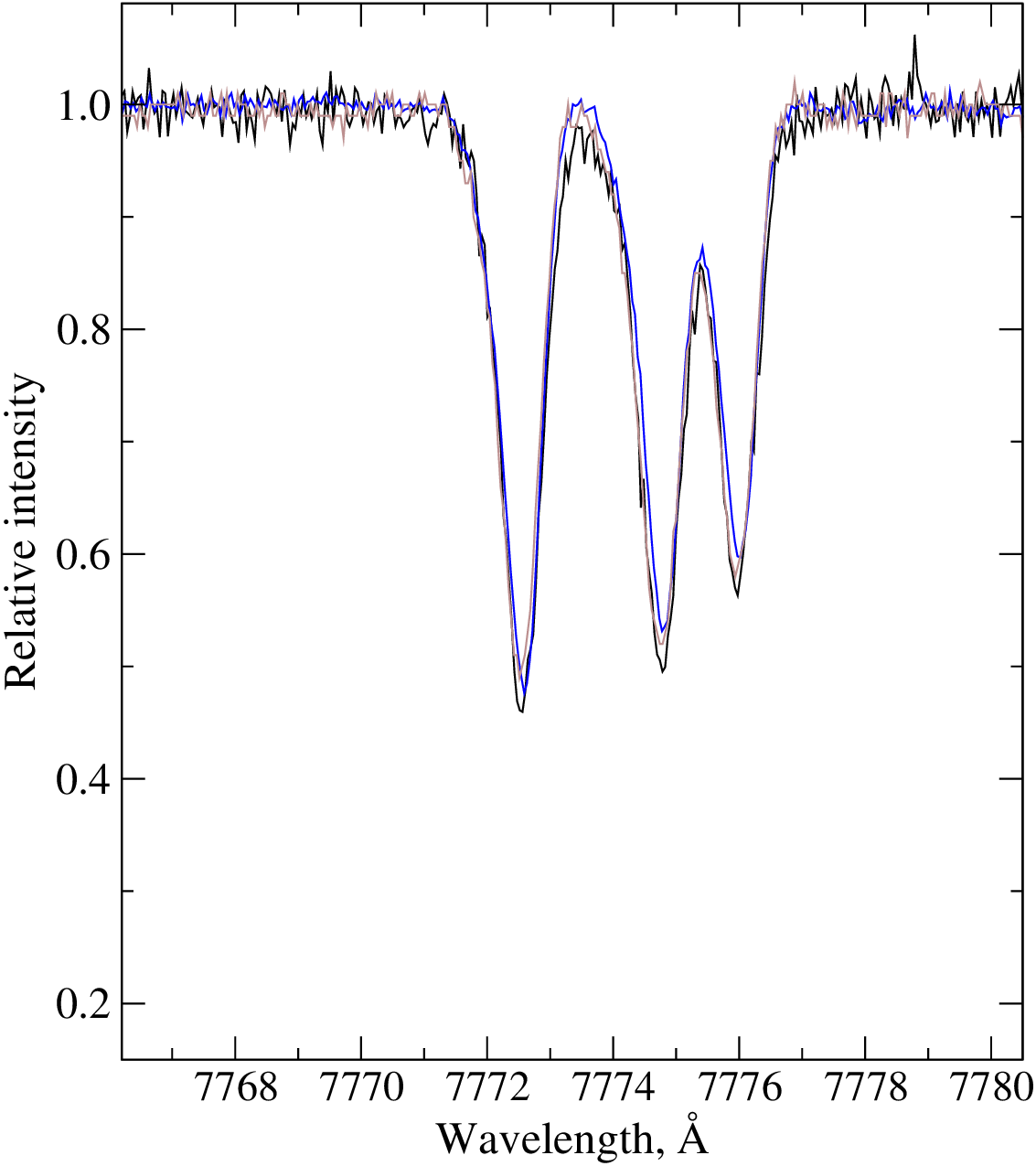}  
\caption{Fragments of PS\,Gem spectra in the region of the O\,I 7773\,\AA\ triplet.
Spectra obtained at orbital phases $\phi = 0.47$ (black line), $\phi = 0.76$ (blue line), and $\phi = 0.19$ (pink line)
are shifted relative to each other along the wavelength axis to compensate for differences in radial velocity.}
\label{O7773}
\end{figure}

The upper panel of Fig.~\ref{CI_477} compares spectra obtained at orbital phases $\phi=0.47$ and $\phi=0.76$. After correcting
for the difference in radial velocity, the strengths and profiles of the C\,I absorptions are nearly identical. This behaviour
indicates that the fundamental atmospheric parameters remain stable throughout the orbital cycle.
Additional evidence for atmospheric stability is provided by the O\,I~7773\,\AA{} triplet shown in
Fig.~\ref{O7773}. The equivalent width measured from all available spectra is remarkably constant, yielding
an average value of $W_{\lambda}(7773)=1.16\pm0.02$~\AA{}.
Using the calibration of \citet{Kovtyukh12}, this equivalent width corresponds to an absolute magnitude of
approximately $M_V\approx-4^{\rm m}$ and a luminosity of  $\log(L/L_{\odot})\approx3.5$.
The luminosity inferred from the empirical calibration of the O\,I\,7773\,\AA\ triplet is consistent with the Gaia-based luminosity estimate of  $L\approx4070\,L_\odot$ reported by \citet{Oudmaijer},  corresponding to $\log(L/L_\odot)\approx3.61$.   The agreement between these two independent estimates supports the post-AGB  evolutionary status of PS\,Gem.

The absence of significant variations in the triplet strength indicates that no substantial luminosity changes
occurred during the observational interval. The derived luminosity is consistent with values typically found for post-AGB stars.

For comparison, the lower panel of Fig.~\ref{CI_477} presents a spectrum of the well-studied post-AGB star CY\,CMi.
Although the atmospheric parameters of CY\,CMi and PS\,Gem are similar, the spectrum of CY\,CMi contains numerous
absorption lines of iron-group and neutron-capture elements that are absent in PS\,Gem. This contrast reflects the
extreme chemical peculiarities produced by depletion processes in the system of PS\,Gem.

The heliocentric radial velocities listed in Table~\ref{tab:obs_rv} were measured using a cross-correlation function
(CCF) technique. The template spectra used in the cross-correlation were synthetic spectra computed with the
\textsc{Synspec} code \cite{Hubeny2021} based on \textsc{ATLAS9} model atmospheres \citep{Kurucz1993}. The adopted
stellar parameters for the template were $T_{\rm eff}=6000$\,K, $\log g=0.5$, ${\rm [Fe/H]}=-4.74$, and a microturbulent
velocity of $v_{\rm mic}=2.0$\,km\,s$^{-1}$. For each observational phase, approximately three to six echelle orders
containing the strongest photospheric absorption lines were selected, while the order including the Na\,D and numerous
telluric lines was excluded because of contamination by strong interstellar absorption. The final radial velocity for each
phase was adopted as the mean value obtained from the selected orders, and the uncertainty was estimated from the scatter
among them. The resulting systemic velocity and semi-amplitude are  $\gamma = 7.46 \pm 0.64$\,km\,s$^{-1}$ and
$K = 2.49$\,km\,s$^{-1}$, respectively.

\subsection{Quantitative Analysis of the H$\alpha$ Variability}

To provide a quantitative characterization of the H$\alpha$ variability, we measured a set of profile parameters for all
spectra covering the H$\alpha$ region. These include the radial velocities of the blue (V) and red (R) emission peaks, the peak
separation $\Delta V_r$, the peak-intensity ratio $I_{\rm V}/I_{\rm R}$, the radial velocity of the absorption core, and
the equivalent width (EW). The measurements are summarized in Table~\ref{tab:Halpha}.

\begin{figure}[hbtp]
\vspace{-2cm}
\includegraphics[width=0.9\linewidth, bb=0 0 500 500,clip]{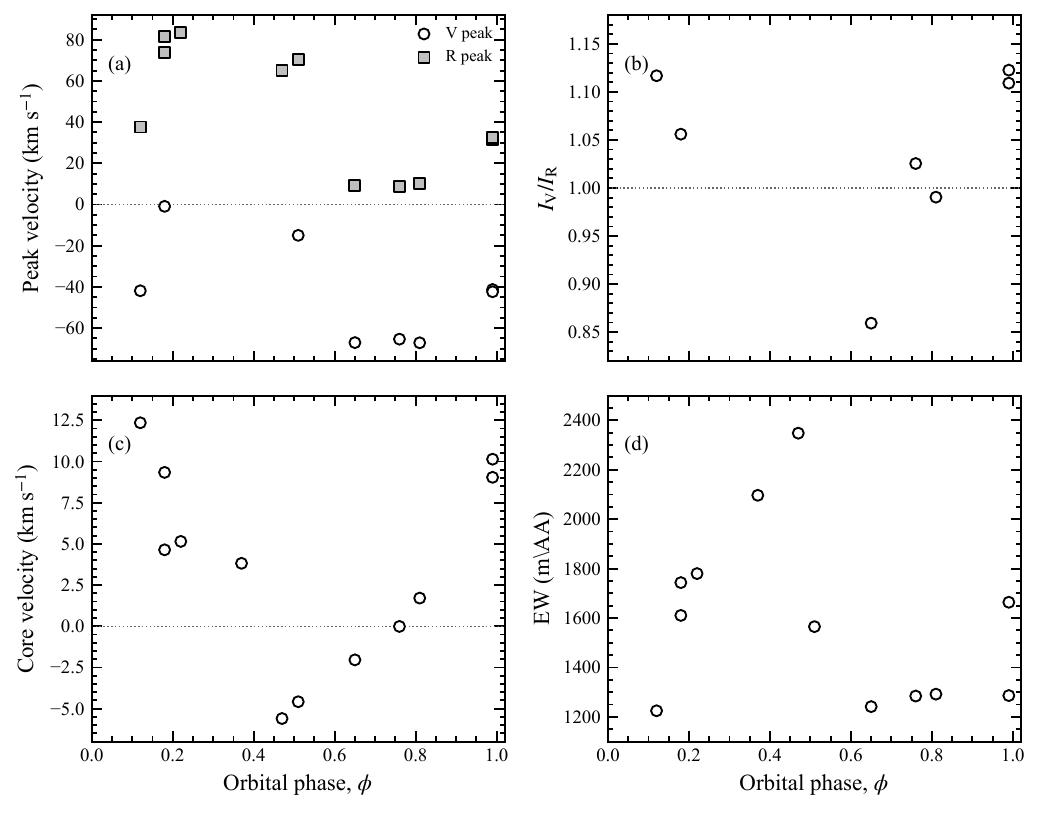}
\centering
\caption{Orbital-phase dependence of the principal H$\alpha$ profile parameters in PS\,Gem.
Panel (a) shows the radial velocities of the blue (V) and red (R) emission peaks.
Panel (b) presents the ratio of the peak intensities,  $I_{\rm V}/I_{\rm R}$.
Panel (c) displays the radial velocity of the H$\alpha$ absorption core.
Panel (d) shows the equivalent width of the H$\alpha$ profile.}
\label{fig:Halpha_phase}
\end{figure}

The H$\alpha$ profiles were normalized to the local continuum before measurement. The radial velocities of the blue and red emission peaks were measured from the wavelengths of the local intensity maxima in the normalized profiles. We did not apply a multi-Gaussian decomposition because the H$\alpha$ profiles are asymmetric and vary significantly with orbital phase. The radial velocity of the absorption core was measured from the minimum of the central absorption component, and the equivalent width was determined by direct numerical integration of the normalized profile.
For spectra in which one or both emission peaks are absent or not reliably resolved, especially near $\phi\approx0.37$ and $\phi\approx0.47$, the corresponding peak velocities and peak separation are not reported. The typical uncertainty of the peak velocities is approximately 1--2\,km\,s$^{-1}$ for well-defined peaks, depending on the signal-to-noise ratio and profile morphology. The quoted value
$\langle\Delta V_r\rangle = 78 \pm 4$\,km\,s$^{-1}$
represents the mean and standard deviation of the measured peak separations among the available spectra, rather than the propagated uncertainty of an individual measurement.

Figure~\ref{fig:Halpha_phase} presents the orbital-phase dependence of the principal H$\alpha$ profile parameters. The radial velocities of the blue and red emission peaks vary systematically with orbital phase (Fig.~\ref{fig:Halpha_phase}a). The peak separation remains within the range of approximately 73--85\,km\,s$^{-1}$, yielding a mean value of
$\langle \Delta V_r \rangle = 78 \pm 4~{\rm km\,s^{-1}}$.
No statistically significant dependence of $\Delta V_r$ on orbital phase is detected.
The peak-intensity ratio $I_{\rm V}/I_{\rm R}$ varies between approximately 0.86 and 1.12 (Fig.~\ref{fig:Halpha_phase}b), indicating significant changes in the relative strengths of the blue and red emission components. The radial velocity of the absorption core ranges from approximately $-5.6$ to $+12.4$\,km\,s$^{-1}$ (Fig.~\ref{fig:Halpha_phase}c), while the equivalent width varies between approximately 1200 and 2400\,m\AA\ (Fig.~\ref{fig:Halpha_phase}d).

The strongest absorption and the largest equivalent widths are observed near $\phi \approx 0.4$, corresponding to epochs when the emission components weaken or disappear. The phase coherence of the measured parameters demonstrates that the H$\alpha$ variability is closely related to the orbital motion of the binary system.

\subsection{Temporal Evolution of the H$\alpha$ Profile}

\begin{figure}[hbtp]
\includegraphics[angle=0,width=0.5\textwidth,bb=0 0 550 675,clip]{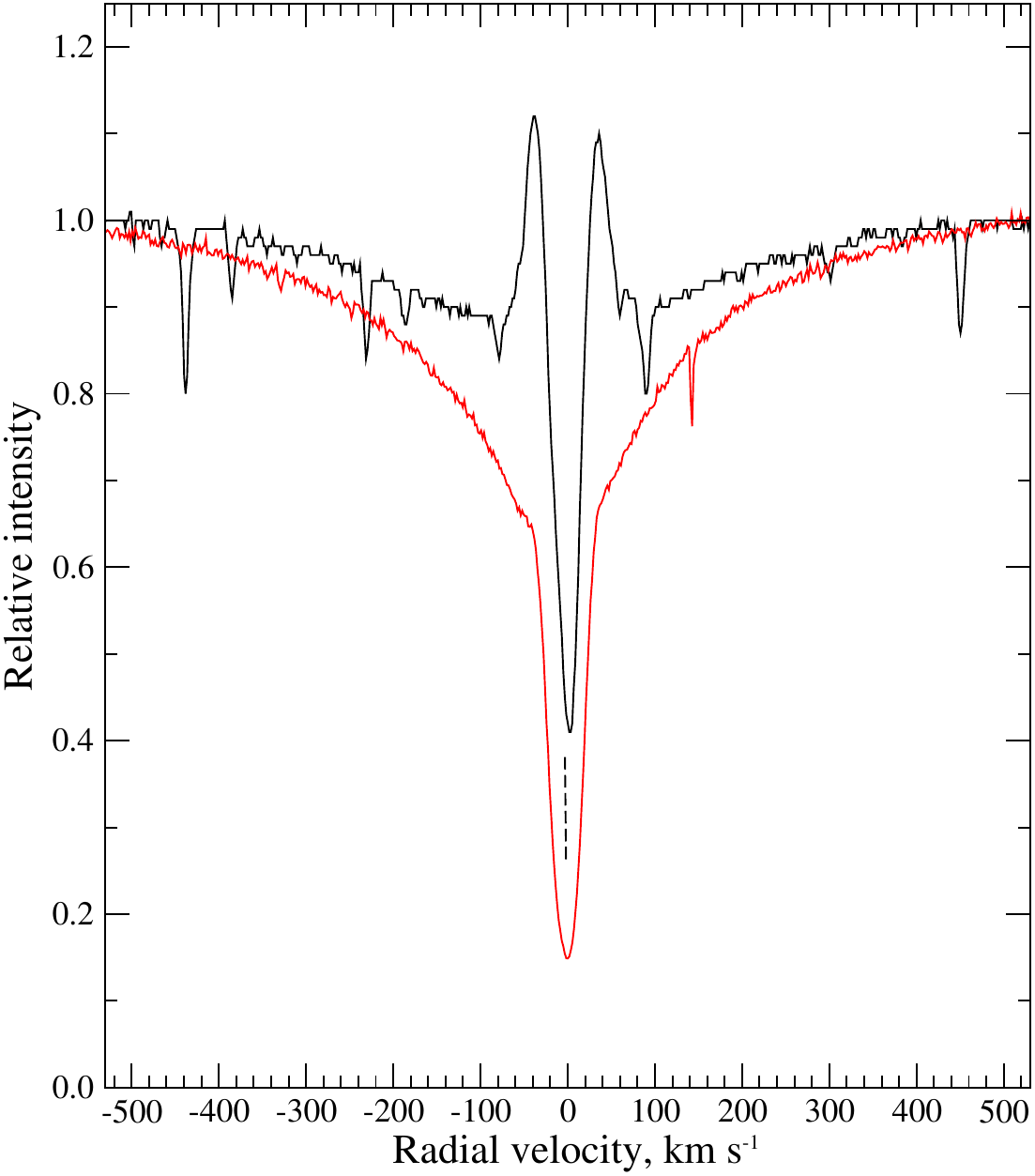}
\caption{The profiles of H$\alpha$ (black line) and H$\beta$ (red line) in the spectrum obtained
     on 11.09.2022 ($\phi$= 0.76). The vertical dotted line marks the heliocentric radial velocity derived from absorption lines, $V_r({\rm abs})$, on the same date. }
\label{Hydrogen}
\end{figure}

The H$\alpha$ line is the only prominent emission feature in the optical spectrum of PS\,Gem. In contrast, the H$\beta$ profile remains purely absorptive (see the Fig.~\ref{Hydrogen}) and is well reproduced by the theoretical profile calculated by \citet{Kipper}.

\begin{figure}[hbtp]
\includegraphics[angle=0,width=0.45\textwidth,bb=0 0 550 675,clip]{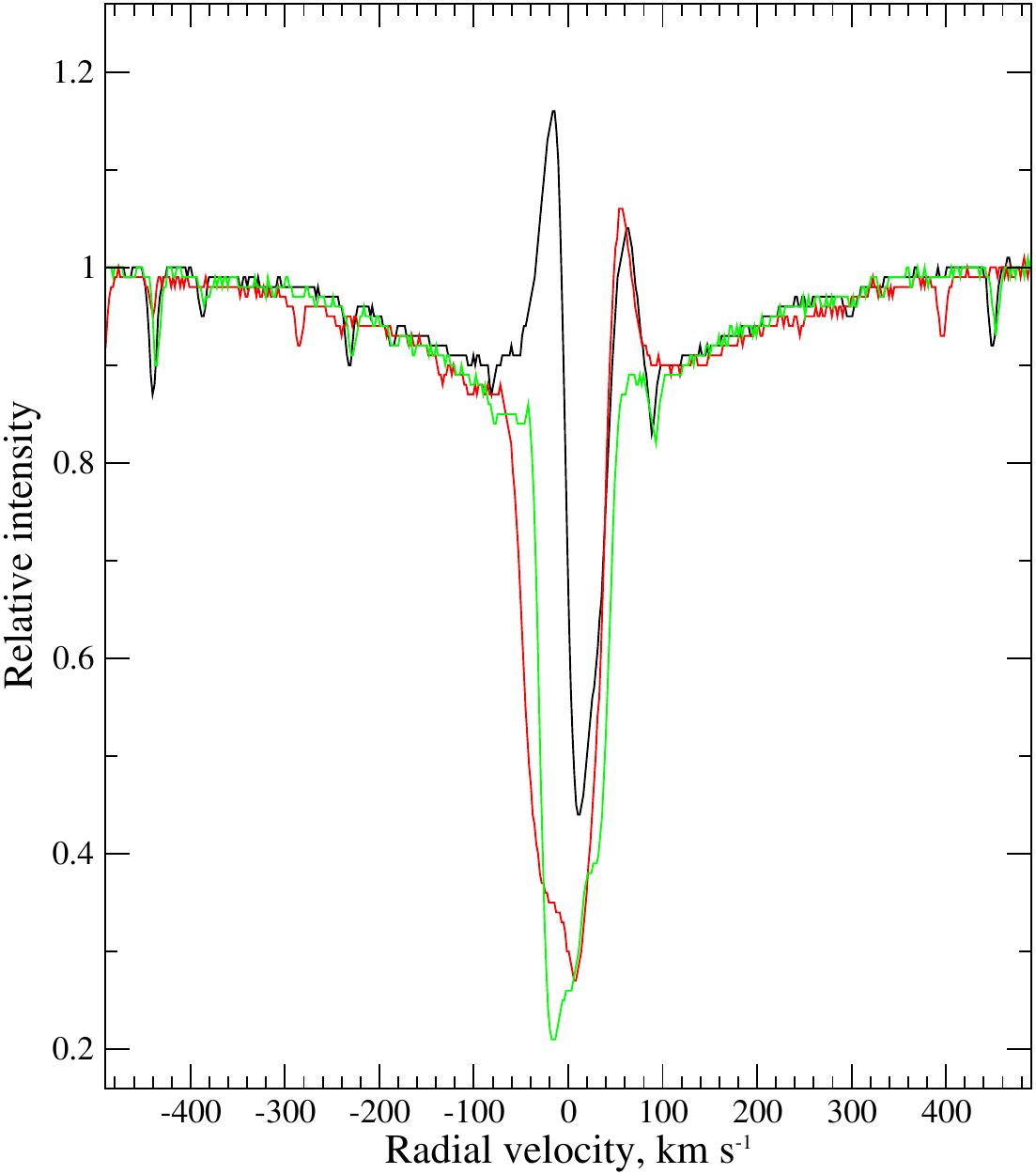}
\includegraphics[angle=0,width=0.45\textwidth,bb=0 0 550 675,clip]{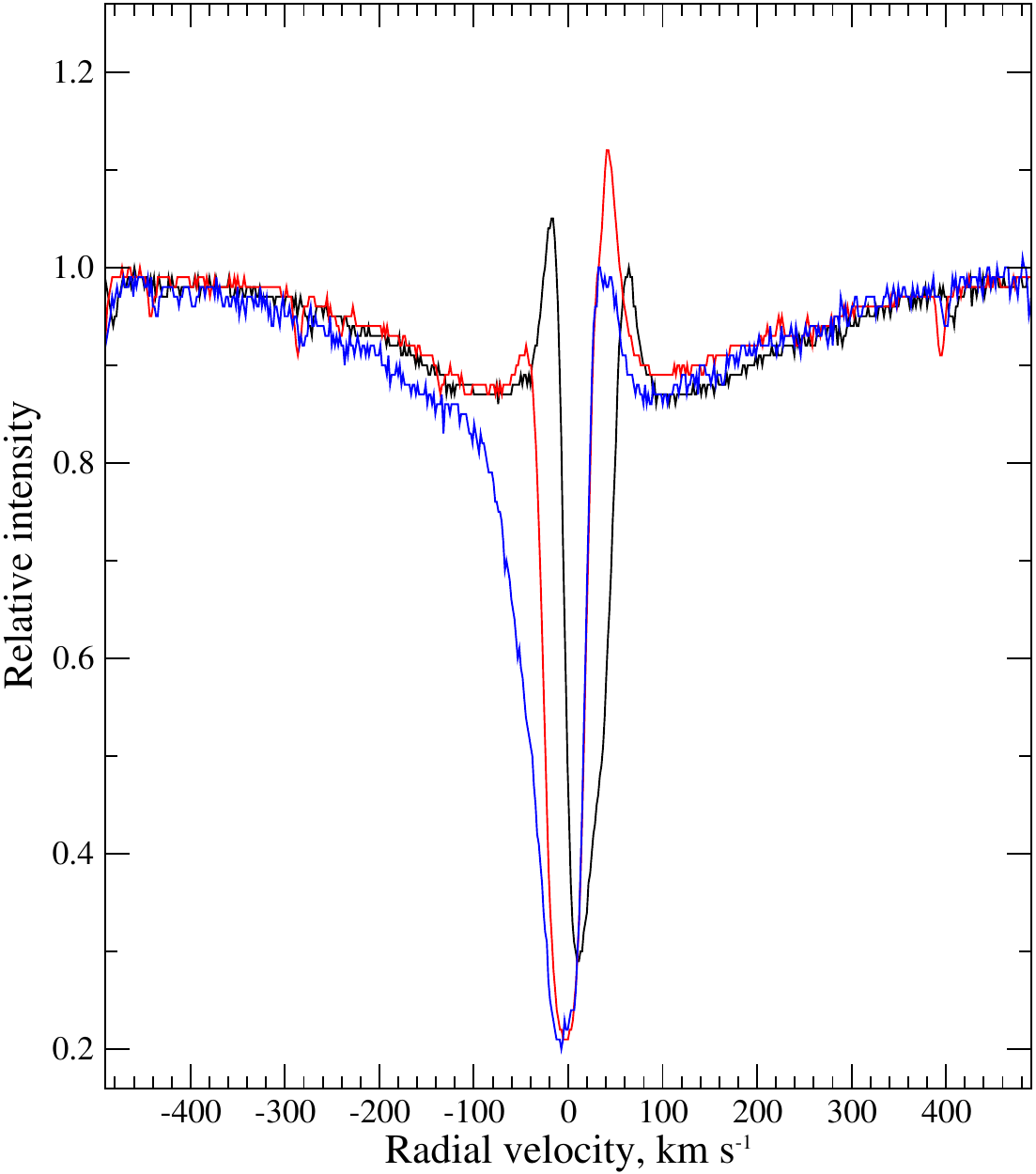}
\caption{Temporal behavior of the H$\alpha$ profile over a time interval of approximately one year.
The left panel shows spectra obtained on 06.11.2009 ($\phi = 0.12$, black line), 02.04.2010 ($\phi = 0.22$, red line), and 25.09.2010 ($\phi = 0.37$, green line).
The right panel presents spectra obtained approximately two orbital cycles later: 13.02.2017 ($\phi = 0.18$, black line), 26.02.2018 ($\phi = 0.47$, blue line), and 05.04.2018 ($\phi = 0.51$, red line).
Telluric absorptions have not been removed in this and the following figures.}
\label{Halpha1}
\end{figure}

\begin{figure}[hbtp]
\includegraphics[angle=0,width=0.5\textwidth,bb=0 0 550 675,clip]{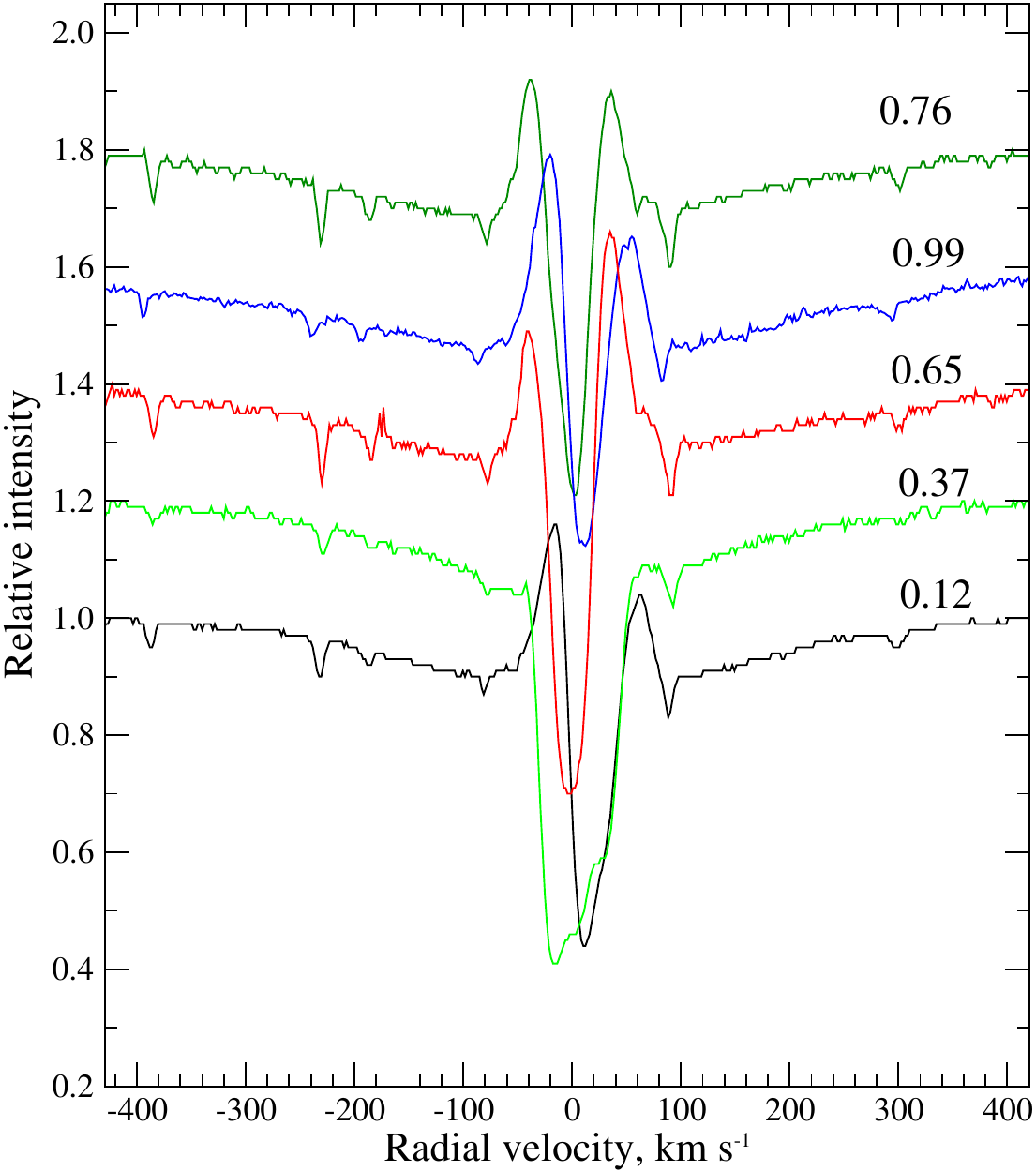}
\caption{H$\alpha$ profiles at different orbital phases in spectra obtained between 2009 and 2022.
The profiles correspond to the following observing dates and phases: 06.11.2009 ($\phi = 0.12$, black line), 25.09.2010 ($\phi = 0.37$, green line), 15.09.2011 ($\phi = 0.65$, red line), 01.12.2012 ($\phi = 0.99$, blue line), and 11.09.2022 ($\phi = 0.76$, dark-green line). The orbital phase is indicated for each profile.}
\label{Halpha2}
\end{figure}

Representative H$\alpha$ profiles obtained at different epochs are displayed in Figs.~\ref{Halpha1} and \ref{Halpha2}. In most spectra, the profile consists of a central absorption component accompanied by weak double-peaked emission wings. No P~Cyg-type profiles are detected, indicating the absence of strong ongoing mass loss.
The temporal evolution of the profile reveals systematic morphological changes over the orbital cycle. Figure~\ref{Halpha1} compares spectra obtained during 2009---2010 with spectra obtained approximately two orbital cycles later. Despite the large temporal separation, the profiles exhibit very similar behaviour at comparable orbital phases, indicating that the variability is reproducible over long timescales.

A particularly remarkable phenomenon occurs near $\phi \approx 0.37$, where the emission components disappear almost completely and the profile becomes dominated by broad absorption. Similar behaviour is observed near $\phi \approx 0.47$, although the emission peaks remain weakly detectable. At these phases, the H$\alpha$ profile resembles those observed in several single post-AGB stars \citep[e.g.][]{V1648Aql,V448Lac,IRAS04296}.

Figure~\ref{Halpha2} illustrates the evolution of the H$\alpha$ profile over the orbital cycle. The profile changes from a configuration with a stronger blue emission peak at early phases to a more symmetric or red-dominated morphology at later phases. Near the completion of the orbital cycle, the profile returns to a shape similar to that observed at earlier phases.
The recurrence of similar profile morphologies at comparable orbital phases, even when separated by several orbital cycles, demonstrates that the H$\alpha$ variability is primarily governed by orbital modulation. Together with the nearly constant velocity scale of the emission peaks, this behaviour suggests that the H$\alpha$ variability is consistent with an origin in a long-lived circumstellar or circumbinary gaseous structure, although this interpretation is not unique.

\subsection{Interstellar and Circumstellar Features}

PS\,Gem is a relatively nearby object with a Gaia EDR3 parallax of $\varpi=0.5178$\,mas and is located only slightly
above the Galactic plane ($b\approx7.6^\circ$). Consistent with its low color excess, $E(B-V)=0.04^{\rm m}$~\cite{Oomen}, only weak interstellar absorption features are expected in the optical spectrum.
Diffuse interstellar bands (DIBs) were identified in ten spectra with adequate signal-to-noise ratios. Between five and ten DIBs were measured in each spectrum, yielding a mean radial velocity $\langle V_r({\rm DIBs}) \rangle = 10.1 \pm 0.8~{\rm km\,s^{-1}}$.
This value agrees well with the radial velocities derived from the interstellar components of the Na\,I\,D and K\,I resonance lines, supporting their common interstellar origin.

\begin{figure}[hbtp]
 \includegraphics[angle=0,width=0.5\textwidth,bb=0 0 550 675,clip]{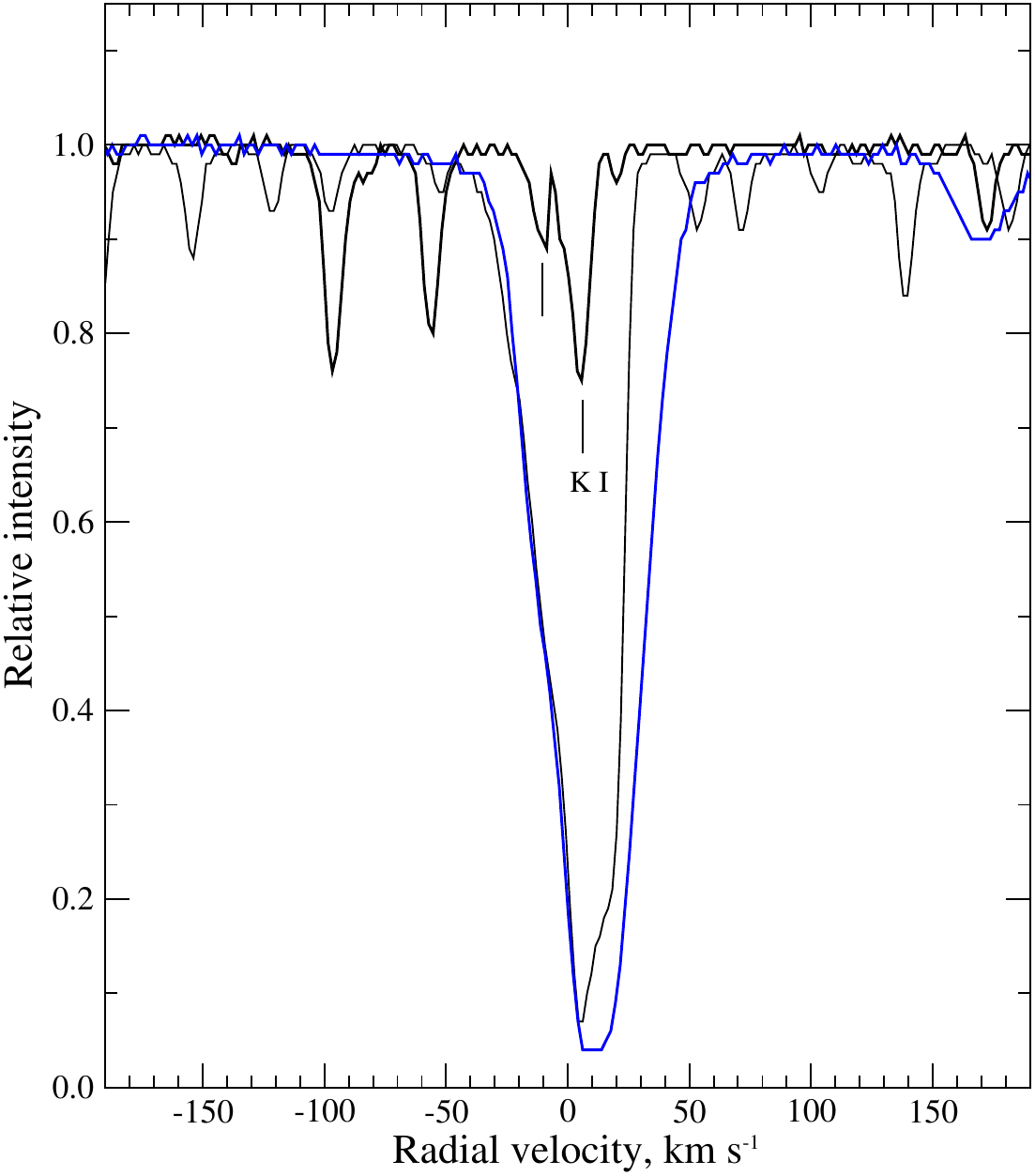}
 \caption{Profiles of the Na\,I 5895\,\AA\ (thin black line), K\,I 7699\,\AA\ (bold black line), and Ca\,II 3933.66\,\AA{} (blue line) features.
The vertical lines indicate the positions of the two K\,I 7699\,\AA\ components at $V_r = +7$ and $-8$\,km\,s$^{-1}$.}
 \label{NaK}
 \end{figure}

Representative profiles of Na\,I\,5895\,\AA\ and K\,I\,7699\,\AA\ observed on 2022 September 11 are shown in Fig.~\ref{NaK}. The Na\,I profile appears as a blend of several unresolved absorptions arising from photospheric, interstellar, and possibly circumstellar contributions. The K\,I\,7699\,\AA\ line shows two absorption components. The redshifted component is close to the interstellar velocity traced by the DIBs and Na\,I lines and is therefore most likely interstellar in origin. The blueshifted component is centered at approximately $V_r\approx -8$\,km\,s$^{-1}$. Using the systemic velocity derived from the homogeneous CCF measurements, $\gamma=7.46\pm0.64$\,km\,s$^{-1}$,
this component is displaced by about 15.5\,km\,s$^{-1}$ with respect to the binary system. If the blueshifted component is associated with circumstellar material, the corresponding projected expansion velocity is therefore of the order of 15--16\,km\,s$^{-1}$. This value is comparable to typical expansion velocities inferred from circumstellar absorption components in evolved and post-AGB stars~\cite{Klochkova2014}. However, the available observations do not allow a unique identification of its formation region. The circumstellar interpretation should therefore be regarded as plausible rather than definitive.

The spectral region containing the Ca\,II H and K lines was recorded in one spectrum obtained on 2010 February 2. Both lines appear as broad, multi-component absorption features. The dominant absorptions are centered near $V_r\approx11$\,km\,s$^{-1}$ and are consistent with the interstellar components identified from the DIBs, Na\,I, and K\,I lines. Additional weak absorptions near the stellar velocity may also be present. However, because PS\,Gem is strongly depleted in calcium, with ${\rm [Ca/H]}=-4.1$ \citep{Kipper}, any photospheric contribution to the Ca\,II profiles is expected to be very weak.

\section{Discussion}\label{discuss}

\subsection{H$\alpha$ Variability and Circumbinary Environment}

The quantitative analysis presented in Section~3 demonstrates that the H$\alpha$ variability of PS\,Gem is closely linked to the orbital motion of the system. The $I_{V}/I_{R}$ ratio and the radial velocity of the absorption core exhibit systematic phase-dependent variations, whereas the separation between the emission peaks remains nearly constant at $\langle \Delta V_r \rangle = 78 \pm 4~{\rm km\,s^{-1}}$.

The absence of a significant phase dependence of $\Delta V_r$ suggests that the characteristic velocity scale of the emitting region remains stable throughout the orbital cycle. The observed peak separation corresponds to a characteristic velocity scale of order
40\,km\,s$^{-1}$. The nearly constant H$\alpha$ peak separation is an observational constraint indicating that the characteristic velocity scale of the emitting region remains stable over the observed phases. A plausible interpretation is that the emission is formed in a long-lived circumstellar or circumbinary gaseous structure whose visibility changes with orbital phase. However, this interpretation is not unique. A detailed physical model of the H$\alpha$-forming region would require substantially denser phase coverage and is beyond the scope of the present work.

In contrast, the $I_{V}/I_{R}$ ratio exhibits clear orbital modulation. Because the velocity separation remains nearly constant while the relative intensities of the peaks vary significantly, the observed changes are more likely related to geometric effects than to global dynamical variations. Possible mechanisms include phase-dependent obscuration by circumstellar material, changes in the visibility of different regions of the emitting structure, or variations in optical depth along the line of sight.

A particularly remarkable phenomenon occurs near $\phi\approx0.37$, where the emission components disappear almost completely and the profile becomes dominated by broad absorption. The available observations do not allow a unique interpretation. However, temporary obscuration of the emitting region, changes in optical depth, or viewing-angle effects associated with the binary configuration may all contribute to this behaviour.

The repeatability of the H$\alpha$ morphology at similar orbital phases over multiple orbital cycles strongly supports an orbital origin of the observed variability. Although the present dataset extends the observational baseline to nearly three decades, it contains only 16 spectra distributed over several orbital cycles. Moreover, some of the spectra do not contain the H$\alpha$ region, therefore, the phase coverage is insufficient for a reliable revision of the orbital solution, and we restrict the present study to a homogeneous characterization of the spectroscopic variability.

\subsection{Evolutionary Status of PS\,Gem}

PS\,Gem belongs to the class of chemically peculiar post-AGB binaries affected by selective depletion. Its extremely low iron abundance, ${\rm [Fe/H]}=-4.8$, together with nearly solar abundances of volatile elements such as C, N, O, S, and Zn, is generally interpreted as the result of gas--dust separation and subsequent reaccretion of metal-poor gas from a circumbinary environment \citep{Waelkens91,Kipper,Oomen}.

The spectroscopic properties of PS\,Gem support this interpretation. The radial-velocity variations confirm its binary nature, while the infrared excess and phase-dependent H$\alpha$ variability indicate the presence of circumstellar or circumbinary material. In contrast to several other depleted post-AGB binaries, PS\,Gem does not exhibit metallic emission lines, P~Cyg-type profiles, or evidence for strong ongoing mass loss. Its optical spectrum is therefore comparatively simple, making it a useful laboratory for studying the interaction between binary evolution and selective depletion.

Among known depleted post-AGB systems, PS\,Gem appears most similar to CC\,Lyr, another extremely metal-poor object showing strong depletion and variable H$\alpha$ emission (\citet{Aoki}). Similar double-peaked H$\alpha$ emission profiles and orbital-phase-dependent  variability have also been reported in  several depleted post-AGB binaries, including HD\,4049 and BD+46\,442
(\citet{Gorlova2012,Gorlova2015}). Both stars exhibit pronounced chemical peculiarities and evidence for a circumbinary environment, yet neither shows clear spectroscopic signatures of the companion. The similarity between these systems suggests that they may represent a distinct subgroup of highly depleted post-AGB binaries.

Despite these similarities, the class of depleted post-AGB binaries remains remarkably diverse. Differences in stellar masses, orbital parameters, disk geometry, viewing angle, and mass-loss history can produce a wide range of photometric and spectroscopic behaviours. In this context, PS\,Gem represents one of the most chemically extreme examples. Its long-term spectroscopic stability, combined with phase-dependent H$\alpha$ variability, provides valuable observational constraints on the role of binary interaction and circumbinary disks in the late stages of stellar evolution.

Although the present dataset spans nearly three decades, the observations are sparsely
distributed over the orbital cycle.  Consequently, the current phase coverage is
insufficient for deriving a revised orbital  solution with improved accuracy.

\section{Conclusions}\label{conclusion}

We have presented the results of long-term high-resolution spectroscopic monitoring of the evolved metal-poor binary PS\,Gem based on 16 spectra obtained with the 6-m BTA telescope during 1997---2025. The principal conclusions of this study are as follows:

\begin{enumerate}

\item
No significant variations are detected  in the photospheric absorption-line spectrum. The equivalent width of the
O\,I 7773\,\AA\ triplet is nearly constant, with an average value of $W_{\lambda}(7773)=1.16\pm0.02$\,\AA,
corresponding to a luminosity $\log(L/L_{\odot})\approx3.5$,  consistent with the post-AGB evolutionary stage of the star.

\item
Radial velocities measured from photospheric absorption lines confirm the binary nature of PS\,Gem. Incorporating the newly recovered spectrum obtained in 2025 yields a mean systemic velocity  $\gamma = 7.46 \pm 0.64$\,km\,s$^{-1}$, with an orbital semi-amplitude of $K = 2.49$\,km\,s$^{-1}$.  The phase dependence of the measured radial velocities agrees well with the orbital solution previously derived by \citet{Winckel1999}.

\item
Weak diffuse interstellar bands are detected in several spectra. Their average radial velocity,
$\langle V_r({\rm DIBs})\rangle = 10.1 \pm 0.8$\,km\,s$^{-1}$,  is consistent with the positions of interstellar
components of the Na\,I and K\,I resonance lines. The K\,I\,7699\,\AA\ profile shows a blueshifted absorption component at
$V_r\approx -8$\,km\,s$^{-1}$.  Relative to the systemic velocity of PS\,Gem,  $\gamma=7.46\pm0.64$\,km\,s$^{-1}$,
this corresponds to a projected velocity offset of about  15.5\,km\,s$^{-1}$.
If this component is circumstellar in origin, the inferred projected expansion velocity is consistent with typical values measured by  \citet{Klochkova2014} for circumstellar envelopes of evolved and post-AGB stars.

\item
The H$\alpha$ profile exhibits pronounced orbital-phase-dependent variability. Quantitative measurements of the profile parameters reveal systematic variations of both the V/R intensity ratio and the radial velocity of the absorption core.
In contrast, the separation between the emission peaks remains nearly constant, with a mean value
$\langle\Delta V_r\rangle = 78 \pm 4$\,km\,s$^{-1}$, indicating a stable velocity field in the emitting region.

\item
The repeatability of the H$\alpha$ morphology at similar orbital phases over multiple orbital cycles demonstrates that
the observed variability is primarily controlled by orbital modulation. The results support the presence of a long-lived circumstellar or circumbinary gaseous structure associated with the binary system.
\end{enumerate}

The nearly three-decade observational baseline presented here provides a valuable foundation for future investigations of PS\,Gem. Additional high-resolution spectra with improved phase coverage will be important for constraining the geometry and kinematics of the circumstellar environment and for developing a more comprehensive picture of the interaction between binary evolution, circumbinary disks, and selective depletion in metal-poor post-AGB systems.

\section*{acknowledgments}
Observations on the BTA telescope of the SAO RAS are carried out with the support of
the Ministry of Science and Higher Education of the Russian Federation. This study is supported
by the National Natural Science Foundation of China (NSFC) under grant No.\,12588202, National Key
R\&D Program of China No.\,2023YFE0107800, No.\,2024YFA1611900, Strategic Priority Research Program
of Chinese Academy of Sciences, grant No.\,1160102.  The study used SIMBAD, SAO/NASA ADS, ASD and
Gaia~EDR3 astronomical databases.


\begin{thebibliography}{99}

\bibitem[Kipper(2013)]{Kipper}
Kipper T., 2013, Baltic Astron.,
\textbf{22}, 101.

\bibitem[Waelkens et al.(1991)]{Waelkens91}
Waelkens C., Van Winckel H., Bogaert E., Trams N.R.,
1991, A\&A, \textbf{251}, 495.

\bibitem[Van Winckel et al.(1999)]{Winckel1999}
Van Winckel H., Waelkens C., Fernie J.D.,
Waters L.B.F.M., 1999, A\&A, \textbf{343}, 202.


\bibitem[Oomen et al.(2018)]{Oomen}
Oomen G.M., Van Winckel H., Pols O., Nelemans G.,
Escorza A., Manick R., Kamath D., Waelkens C.,
2018, A\&A, \textbf{620}, A85.


\bibitem[Panchuk et al.(1997)]{PFES}
Panchuk V.E., Najdenov I.D., Klochkova V.G.,
Ivanchik A.V., Yermakov S.V., Murzin V.A.,
1997, Bull. Spec. Astrophys. Obs., \textbf{44}, 127.

\bibitem[Panchuk et al.(2017)]{NES}
Panchuk V.E., Klochkova V.G., Yushkin M.V.,
2017, Astron. Rep., \textbf{61}, 820.


\bibitem[Yushkin \& Klochkova(2004)]{echelle}
Yushkin M.V., Klochkova V.G., 2004, SAO Preprint, No.206.

\bibitem[Galazutdinov(2022)]{dech}
Galazutdinov G.A., 2022, Astrophys. Bull., \textbf{77}, 519.


\bibitem[Klochkova et al.(2007)]{atlas}
Klochkova V.G., Chentsov E.L., Tavolganskaya N.S.,
Shapovalov M.V., 2007, Astron. Rep., \textbf{94}, 162.

\bibitem[Kovtyukh et al.(2012)]{Kovtyukh12}
Kovtyukh V.V., Gorlova N.I., Belik S.I.,
2012, MNRAS, \textbf{423}, 3268.


\bibitem[Oudmaijer et al.(2022)]{Oudmaijer}
Oudmaijer R.D., Jones E.R.M., Vioque M.,
2022, MNRAS, \textbf{516}, L61.



\bibitem[Hubeny et al.(2021)]{Hubeny2021}
Hubeny I., Allende Prieto C., Osorio Y., et al.,
2021, arXiv:2104.02829.


\bibitem[Kurucz(1993)]{Kurucz1993}
Kurucz R., 1993, Robert Kurucz CD-ROM, 13.


\bibitem[Klochkova \& Tavolzhanskaya(2019)]{V1648Aql}
Klochkova V.G., Tavolzhanskaya N.S., 2019,
Astrophys. Bull., \textbf{74}, 277.

\bibitem[Klochkova et al.(2010)]{V448Lac}
Klochkova V.G., Panchuk V.E., Tavolzhanskaya N.S.,
2010, Astron. Rep., \textbf{54}, 234.

\bibitem[Klochkova et al.(1999)]{IRAS04296}
Klochkova V.G., Szczerba R., Panchuk V.E., Volk K.,
1999, A\&A, \textbf{345}, 905.


\bibitem[Klochkova(2014)]{Klochkova2014}
Klochkova V.G., 2014, Astrophys. Bull., \textbf{69}, 279.

\bibitem[Aoki et al.(2017)]{Aoki}
Aoki W., Matsuno T., Honda S., Parthasarathy M.,
Li H., Suda T., 2017, PASJ, \textbf{69}, 21.


\bibitem[Gorlova et al.(2012)]{Gorlova2012}
Gorlova N., Van Winckel H., Gielen C., Raskin G.,
Prins S., Pessemier S.W., Waelkens C., Fremat Y.,
Hensberge H., Dumortier L., Jorissen A., Van Eck S.,
2012, A\&A, \textbf{542}, A27.

\bibitem[Gorlova et al.(2015)]{Gorlova2015}
Gorlova N., Van Winckel H., Ikonnikova N.P.,
Burlak M.A., Komissarova G.V., Jorissen A.,
Gielen I.C., Debosscher J., Degroote P.,
2015, MNRAS, \textbf{451}, 2462.

\end{thebibliography}
\end{document}